\documentstyle[preprint,aps]{revtex}
\tightenlines
\draft

\newcommand{\be}{\begin{equation}}
\newcommand{\ee}{\end{equation}}
\newcommand{\beal}{\begin{eqalign}}
\newcommand{\eeal}{\end{eqalign}}
\newcommand{\bea}{\begin{eqnarray}}
\newcommand{\eea}{\end{eqnarray}}
\newcommand{\bean}{\begin{eqnarray*}}
\newcommand{\eean}{\end{eqnarray*}}
\newcommand{\ba}{\begin{array}}
\newcommand{\ea}{\end{array}}

\newcommand{\ka}{\kappa}
\newcommand{\La}{\Lambda}
\newcommand{\la}{\lambda}

\newcommand{\Om}{\Omega}
\newcommand{\de}{\delta}

\newcommand{\pa}{\partial}

\newcommand{\no}{\nonumber}

\newcommand{\tr}{\mbox{tr}}
\newcommand{\res}{\mbox{res}}

\begin{document}

\title
 { \sc On the B\"acklund Transformation for the\\
  Moyal Korteweg-de Vries Hierarchy\/}
\author{
{\sc Ming-Hsien Tu\footnote{E-mail: phymhtu@ccunix.ccu.edu.tw}\/}\\
  {\it Department of Physics, National Chung Cheng University,\\
   Minghsiung, Chiayi, Taiwan\/}\\
   }
\date{\today}
\maketitle
\begin{abstract}
 We study the B\"acklund symmetry for the Moyal Korteweg-de Vries (KdV) hierarchy
 based on the Kuperschmidt-Wilson Theorem associated with second Gelfand-Dickey structure
  with respect to the Moyal bracket, which generalizes the result of Adler for the
   ordinary KdV.
   \end{abstract}

\newpage
%%%%%%%%%%%%%%%%%%%%%%%%%%%%%%%%%%%%%%%%%%%%%%%%%%%%%%%%%%%%%%%%%%%%
In \cite{A}, Adler provided an elegant approach to the B\"acklund transformation for
 the generalized Korteweg-de Vries (KdV) equations \cite{GD,D} by exploring the hidden
  symmetry of the associated modified KdV equations. The approach is intimately related
  to the Kupershnidt-Wilson (KW) theorem \cite{KW} (see also \cite{D2,MR}) for the second
 Gelfand-Dickey (GD) bracket \cite{GD,D}, the Hamiltonian structure of the KdV, in which
  the Miura variables play an important role. Based upon which, some generalizations
  including Toda \cite{A}, Drinfeld-Sokolov \cite{GRUW}, KP \cite{GU,C,D3}, supersymmetric
   KdV \cite{ST} and $q$-deformed KdV\cite{TL} hierarchies have been discussed in a similar way.

In this Letter, we would like to generalize the approach to the second
GD structure with respect to the Moyal bracket \cite{M} defined by
\be
 \{f,g\}_\ka=\frac{f\star g-g\star f}{2\ka}
 \label{Moyalb}
\ee
 where $f$ and $g$ are two arbitrary functions on the two-dimensional phase space with
 coordinate $(x,p)$ and the $\star$-product \cite{Gr} is defined by
 \be
f\star g=\sum_{s=0}^{\infty}\frac{\ka^s}{s!}\sum_{j=0}^s(-1)^j{s \choose j}(\pa_x^j\pa_p^{s-j}f)
 (\pa_x^{s-j}\pa_p^jg)
 \label{Moyal2}
\ee
The Moyal bracket (\ref{Moyalb}) can be viewed as the higher-order derivative
  generalization of the canonical Poisson bracket since it recovers the canonical Poisson
  bracket in the dispersionless limit $\ka\to 0$, i.e.
  $\lim_{\ka \to 0}\{f,g\}_\ka=\pa_pf\pa_xg-\pa_xf\pa_pg$. Therefore the Lax equations defined
  by the Moyal bracket can be viewed as a one-parameter deformation of dispersionless Lax
  equations (see \cite{TT} for a review) defined by the canonical Poisson bracket.

Our  main results contain two parts. In Theorem A, we prove the KW theorem for the second
GD structure with respect to the Moyal bracket, which simplifies the associated
 hamiltonian flows in terms of the Miura variables. We then, in Theorem B, show that
  the B\"acklund transformation for the generalized Moyal KdV hierarchy can be traced back
  to the permutation symmetry of the Miura variables via the KW theorem and thus
   generalize Adler's work to the present case.

 To begin with, we consider an algebra of Laurent series of the form
 $\La=\{A|A=\sum_{i=-\infty}^Na_ip^i\}$ with coefficients $a_i$ depending on an
  infinite set of variables $t_1\equiv x, t_2, t_3, \cdots$. The algebra $\La$ with
  respect to the Moyal bracket (\ref{Moyalb})  can be decomposed into the sub-algebras
   as $\La_+\oplus \La_-$ where the subscript $+$ stands for the projection onto the
   non-negative powers in $p$. It's obvious that $\La$ is an  associative but noncommutative
    algebra under the $\star$-product. For a given Laurent series $A=\sum_ia_ip^i$ one
     defines its  residue as $\res(A)=a_{-1}$  and its trace as $\tr(A)=\int \res(A)$.
For any two Laurent series $A=\sum_{i}a_ip^i$ and $B=\sum_{j}b_jp^{-j}$ it is easy to show that
(i) $\int\res(A\star B)=\sum_i\int a_ib_{i+1}$ (ii) $\tr \{A, B\}_\ka=0$
(iii) $ \tr (A\star\{B, C\}_\ka)=\tr (\{A, B\}_\ka\star C)$.
 Given a functional $F(A)=\int f(a)$ we define its gradient as
 $\de F/\de A\equiv d_AF=\sum_i \de f/\de a_ip^{-i-1}$ where the variational derivative
  is defined by $\de f/\de a_k=\sum_i(-1)^i(\pa^i\cdot\pa f/\pa a_k^{(i)})$,
with $a_k^{(i)}\equiv (\pa^i\cdot a_k), \pa\equiv \pa/\pa x$.
 Note that we use the notations $\pa_xf=f'=f_x$
and $\pa f=f\pa+f'$ throughout this paper.

Let us consider the Lax equations
 \be
  \frac{\pa L}{\pa t_k}=\{(L^{1/n}\star)^k_+,L\}_{\ka},\qquad
 (L^{1/n}\star)^k_+=(\underbrace{L^{1/n}\star L^{1/n}\star\cdots \star L^{1/n}}_{k})_+
  \label{laxeq}
\ee
 where the Lax operator $L=p^n+\sum_{i=0}^{n-2}u_ip^{i}$ is a polynomial in $p$ and
  $L^{1/n}=p+\sum_{i=1}a_ip^{-i}$ is the $n$th root of $L$ in such a way that
$L=(L^{1/n}\star)^n.$ The Lax equation (\ref{laxeq}) define the dynamical flows
for $u_i$, which form what we call the generalized Moyal KdV hierarchy. Note that
 the highest order in $p$ on the right-hand side  of the Lax equations (\ref{laxeq}) is $n-2$,
   and thus one can drop the term  $u_{n-1}$ in the Lax formulation without
   causing any problem. However, we shall see that imposing the constraint $u_{n-1}=0$
   induces a modification in the Hamiltonian formulation.

For the polynomial $L=p^n+\sum_{i=0}^{n-1}u_ip^i$ and functionals $F[L]$ and $G[L]$
 we define the second GD bracket \cite{GD,D} with respect to the
  $\star$-product as
  \be
\{F,G\}_2(L)=\tr\left[J_2(d_LF)\star d_LG\right]
 \label{gdb}
  \ee
 in which $J_2$ is the Adler map \cite{A2} defined by
   \be
  J_2(X)=\{L,X\}_{\ka+}\star L-\{L,(X\star L)_+\}_\ka,
    \label{adler}
  \ee
  where $X=\sum_{i=1}^nx_ip^{-i-1}$. The bracket (\ref{gdb}) is anti-symmetric due to the
   cyclic property of the trace and satisfies the Jacobi identity
  that will be justified later on. In the $\ka\to 0$ limit (\ref{gdb}) recovers the
   dispersionless GD bracket \cite{FR}.

{\sc Theorem A\/} (Kupershmidt and Wilson). Let $L=A\star B$ where both $A$ and $B$ are
 polynomials in $p$, then we have
\be
\{F,G\}_2(L)=\{F,G\}_2(A)+\{F,G\}_2(B).
\label{fact}
\ee
{\it Proof.\/} From the variation
\bean
\de F&=&\tr(d_LF\star \de L)\\
&=&\tr(d_LF\star \de A\star B+d_LF\star A\star \de B)\\
&=&\tr(d_AF\star \de A)+\tr(d_BF\star \de B)
\eean
we have $d_AF=B\star d_LF$ and $ d_BF=d_LF\star A$. Then
\bean
R.H.S.&=&\tr[\{A,d_AF\}_{\ka+}\star A\star d_AG-\{A, (d_AF\star A)_+\}_\ka\star d_AG]+
(A\leftrightarrow B)\\
&=&\frac{1}{2\ka}\tr[(A\star d_AF)_+\star A\star d_AG-(d_AF\star A)_+\star A\star d_AG)\\
&&-A\star (d_AF\star A)_+\star d_AG+(d_AF\star A)_+\star A\star d_AG]+(A\leftrightarrow B)\\
&=&\frac{1}{2\ka}\tr[(A\star d_AF)_+\star A\star d_AG-A\star (d_AF\star A)_+\star d_AG]
+(A\leftrightarrow B)\\
&=&\tr[J_2(d_LF)\star d_LG].\qquad \Box
\eean
 If we factorize the Lax operator as
 \be
 L=p^n+\sum_{i=0}^{n-1}u_ip^i=
 (p-\phi_n)\star\cdots \star(p-\phi_1)
 \label{factl}
 \ee
then the coefficients $u_i$ can be expressed in terms of Miura variables $\phi_i$ as
\bea
u_{n-1}&=&-\sum_{i=1}^n\phi_i,\no\\
u_{n-2}&=&\sum_{i>j}\phi_i\phi_j-\ka\sum_{i=1}^n(n-2i+1)\phi_i'\label{mt},\\
&\vdots& \no
\eea
which constitute the Miura transformation. Therefore under the factorization
 (\ref{factl}), by Theorem A, the second GD structure (\ref{gdb}) can be simplified as follows
\[
\{F,G\}_2(L)=\sum_{i=1}^n\int \left(\frac{\de F}{\de \phi_i}\right)'
\left(\frac{\de G}{\de \phi_i}\right).
\]
In particular,
\be
\{\phi_i(x), \phi_j(y)\}_2=-\de_{ij}\pa_x\de(x-y)
\label{free}
\ee
which immediately verifies the Jacobi identity of the GD bracket (\ref{gdb}).

The second Hamiltonian structure for the generalized Moyal KdV hierarchy
can be obtained from (\ref{gdb}) by imposing the constraint $u_{n-1}=-\sum_{i=1}^n\phi_i=0$.
Using (\ref{free}) we obtain $\{u_{n-1}(x),u_{n-1}(y)\}_2=-n\pa_x\de(x-y)$ which has an inverse
for $n\neq 0$ and hence the constraint is second class.
The standard Dirac procedure thus gives the modified Adler map as
\be
J_2^D(X)=\{L,X\}_{\ka+}\star L-\{L,(X\star L)_+\}_\ka+
\frac{1}{n}\left\{L,\int^x\res\{L,X\}_\ka\right\}_\ka
\label{madler}
\ee
or, in terms of Miura variables,
\be
\{\phi_i(x), \phi_j(y)\}^D_2=\left[\frac{1}{n}-\de_{ij}\right]\pa_x\de(x-y).
\label{dmiura}
\ee
On the other hand, one might obtain the associated first GD structure
 for the Moyal KdV by  shifting $L\to L+\la$ in (\ref{madler}) and then extract
  the term linear in $\la$ \cite{D}.  It turns out that
  \be
J_1(X)=\{L,X\}_{\ka+}
\label{adler1}
  \ee
which is compatible with the reduction $u_{n-1}=0$.
 As a result, the bi-Hamiltonian flows for the generalized Moyal KdV hierarchy can be written as
\be
\frac{\pa L}{\pa t_k}=J_2^D(d_LH_k)=J_1(d_LH_{k+n})
\label{biham}
\ee
with
\[
H_k=\frac{n}{k}\int\res(L^{1/n}\star)^k.
\]

Next we turn  to the Ba\"cklund transformation for the Moyal KdV hierarchy. Our strategy
is to rewrite the hierarchy flows in terms of the Miura variables. Following Adler \cite{A}
we define the permutation  $\Om: \phi_1\to\phi_2, \phi_2\to\phi_3, \cdots,
\phi_n\to\phi_1$, then $L_{\Om^i}=(p-\phi_i)\star (p-\phi_{i-1})\star\cdots\star (p-\phi_{i+1})$.
In particular, $L_{\Om^n}=L_{\Om^0}=L$ and $L_{\Om^i}=p_i\star L_{\Om^{i-1}}\star p_i^{-1} $ where
$p_i\equiv p-\phi_i$ and its inverse $p_i^{-1}$ can be expressed as
$\exp(\int^x\phi_i/2\ka)\star p\star\exp(-\int^x\phi_i/2\ka)$ and
 $\exp(\int^x\phi_i/2\ka)\star p^{-1}\star\exp(-\int^x\phi_i/2\ka)$, respectively.

{\sc Theorem B\/} (Adler). The Miura variables $\phi_i$ satisfy the follwoing modified
Moyal KdV equations
\be
\frac{\pa \phi_i}{\pa t_k}=\frac{1}{2\ka}\left[p_i\star B_{\Om^{i-1}}^{(k)}-
B_{\Om^{i}}^{(k)}\star p_i\right]
\label{thm}
\ee
where $B_{\Om^i}^{(k)}\equiv(L_{\Om^i}^{1/n}\star)^k_+$.\\
Before proving (\ref{thm}) we observe that (\ref{thm}) is compatible with the Lax flows
(\ref{laxeq}) since from the factorization form of $L$ we have
\bean
\frac{\pa L}{\pa t_k}&=&-\sum_{i=1}^np_n\star\cdots\star p_{i+1}\star\frac{\pa \phi_i}{\pa t_k}
\star p_{i-1}\star\cdots\star p_1,\\
&=&-\frac{1}{2\ka}\sum_{i=1}^np_n\star\cdots\star p_{i+1}\star
\left(p_i\star B_{\Om^{i-1}}^{(k)}-B_{\Om^{i}}^{(k)}\star p_i\right)\star p_{i-1}
\star\cdots\star p_1,\\
&=&\{B^{(k)}, L\}_{\ka}.
\eean

{\sc Lemma.\/} The Hamiltonian flows for the Miura variables $\phi_i$ can be expressed as
\be
\frac{\pa \phi_i}{\pa t_k}=\res\left\{p_i, p_{i-1}\star\cdots\star p_1\star
 (L^{1/n}\star)^{k-n}\star p_n\star\cdots\star p_{i+1}\right\}_\ka.
\label{lemma}
\ee
{\it Proof\/}. From the variation
\bean
\frac{\de L}{\de \phi_i}&=&\frac{\de L^{1/n}}{\de\phi_i}\star L^{1/n}\star\cdots\star L^{1/n}\\
&&+L^{1/n}\star\frac{\de L^{1/n}}{\de\phi_i}\star\cdots\star L^{1/n}\\
&&\vdots\\
&&+L^{1/n}\star L^{1/n}\star\cdots\star\frac{\de L^{1/n}}{\de\phi_i}
\eean
we have
\bean
\frac{\de H_k}{\de\phi_i}&=&n\tr\left[\frac{\de L^{1/n}}{\de\phi_i}
\star (L^{1/n}\star)^{k-1}\right],\\
&=&\tr \left[\frac{\de L}{\de \phi_i}\star (L^{1/n}\star)^{k-n}\right],\\
&=&-\res\left[p_{i-1}\star\cdots\star p_1\star (L^{1/n}\star)^{k-n}\star p_n
\star\cdots\star p_{i+1}\right].
\eean
Hence
\bean
\frac{\pa \phi_i}{\pa t_k}&=&\{\phi_i, H_k\}_2^D=\sum_j\left[\frac{1}{n}-\de_{ij}\right]
\left(\frac{\de H_k}{\de \phi_j}\right)',\\
&=&-\sum_j\left[\frac{1}{n}-\de_{ij}\right]\left\{p_j, \res\left[p_{j-1}\star\cdots\star p_1\star
 (L^{1/n}\star)^{k-n}\star p_n\star\cdots\star p_{j+1}\right]\right\}_\ka,\\
&=&\res\left\{p_i, p_{i-1}\star\cdots\star p_1\star (L^{1/n}\star)^{k-n}\star p_n
\star\cdots\star p_{i+1}\right\}_\ka.\qquad \Box
\eean
{\it Proof of Theorem B\/}. From the relation $L_{\Om^i}=p_i\star L_{\Om^{i-1}}\star p_i^{-1}$
we have
\bean
B_{\Om^i}^{(k)}&=&\left(L_{\Om^i}^{1/n}\star\right)^k_+,\\
&=&\left(p_i\star B_{\Om^{i-1}}^{(k)}\star p_i^{-1}\right)_+,\\
&=&p_i\star B_{\Om^{i-1}}^{(k)}\star p_i^{-1}-
\left(p_i\star B_{\Om^{i-1}}^{(k)}\star p_i^{-1}\right)_-.
\eean
Then
\bean
\frac{1}{2\ka}\left[p_i\star B_{\Om^{i-1}}^{(k)}-B_{\Om^{i}}^{(k)}\star p_i\right]
&=&\frac{1}{2\ka}\left(p_i\star B_{\Om^{i-1}}^{(k)}\star p_i^{-1}\right)_-\star p_i,\\
&=&\frac{1}{2\ka}\res\left(p_i\star B_{\Om^{i-1}}^{(k)}\star p_i^{-1}\right),\\
&=&\frac{1}{2\ka}\left[\res\left(p_i\star (L_{\Om^{i-1}}^{1/n}\star)^k\star p_i^{-1}\right)
-\res\left(L_{\Om^{i-1}}^{1/n}\star\right)^k\right]\\
&=&\res\left\{p_i, p_{i-1}\star\cdots\star p_1\star (L^{1/n}\star)^{k-n}\star p_n
\star\cdots\star p_{i+1}\right\}_\ka
\eean
which,  by {\sc Lemma\/}, completes the proof of Theorem B. $\Box$\\
We emphasize here that (\ref{lemma}) (and hence (\ref{thm})) is consistent with the constraint
 $u_{n-1}=-\sum_i\phi_i=0$.

Thus the cyclic permutation $\Om$ generates the B\"acklund transformation
 of the hierarchy due to the fact that the form of the Lax operator and the hierarchy
  flows are preserved under such transformation.
  In particular, the one-step permutation $\Om$: $L_{\Om^{i-1}}\to L_{\Om^i}$
  defines an elementary B\"acklund transformation. Indeed, if $L_{\Om^{i-1}}$ satisfies
   (\ref{laxeq}), then the transformed Lax operator $L_{\Om^i}$ satisfies
  \bean
\frac{\pa L_{\Om^i}}{\pa t_k}&=&\left\{p_i\star B^{(k)}_{\Om^{i-1}}\star p_i^{-1}+2\ka
\frac{\pa p_i}{\pa t_k}\star p_i^{-1}, L_{\Om^i}\right\}_\ka\\
&=&\left\{p_i\star B^{(k)}_{\Om^{i-1}}\star p_i^{-1}-2\ka
\frac{\pa \phi_i}{\pa t_k}\star p_i^{-1},L_{\Om^i}\right\}_\ka\\
&=&\left\{B^{(k)}_{\Om^i},L_{\Om^i}\right\}_\ka
 \eean
 where (\ref{thm}) was used to reach the last line.

Finally let us work out the simplest example to illustrate the obtained results.
 For $n=2$ we have $L=p^2+u$ and the first few Lax flows are given by
 \bea
 u_{t_1}&=&u_x,\no\\
 u_{t_3}&=&\frac{3}{2}uu_x+\ka^2u_{xxx},\no\\
 u_{t_5}&=&\frac{15}{8}u^2u_x+\frac{5}{2}\ka^2(uu_{xxx}+2u_xu_{xx})+\ka^4u^{(5)},
 \label{Moyalkdv}\\
 &\vdots&\no
  \eea
 The set of equations (\ref{Moyalkdv}) is referred to as the Moyal KdV hierarchy
  \cite{Ku,St,Ko,Tu,DP}. For the Hamiltonian formulation,
  we can read off the Poisson brackets from (\ref{biham}) as \cite{Tu,DP}
 \bea
\{u(x),u(y)\}_1&=&2\pa_x\delta(x-y),\no\\
 \{u(x),u(y)\}^D_2&=&[2\ka^2\pa_x^3+2u\pa_x+u_x]\delta(x-y).
 \label{kdvb}
 \eea
 The first few Hamiltonians for the Moyal KdV are
 \bean
 H_1&=& \int u, \quad H_3= \frac{1}{4}\int u^2,\quad
 H_5= \frac{1}{8}\int (u^3+2\ka^2 uu_{xx}),\\
 H_7&=& \frac{1}{64}\int (5u^4-40\ka^2 uu_x^2+16\ka^4u_{xx}^2),
\eean
 which together with (\ref{kdvb}) provides the bi-Hamiltonian flows
 \[
 \frac{\pa u}{\pa t_{2n+1}}=[2\ka^2\pa_x^3+2u\pa_x+u_x] \frac{\de H_{2n+1}}{\de u}
 =2\pa_x\frac{\de H_{2n+3}}{\de u}.
 \]
We remark that when $\ka=0$ the Moyal KdV hierarchy reduces to the dispersionless KdV hierarchy
\cite{Kr} since all higher-order derivative  terms disappear. One the other hand,
 (\ref{Moyalkdv}) collapses to the ordinary KdV for $\ka=1/2$ due to an isomorphism between them
 \cite{G} and thus the Moyal parameter $\ka$ characterizes a kind of dispersion effect.

Furthermore consider the factorization of the Lax operator $L=p^2+u=(p-\phi)\star(p+\phi)$
which gives the Miura transformation (or Riccati relation) $u(\phi)=-\phi^2+2\ka\phi'$.
Then the Poisson algebra (\ref{kdvb}) can be easily rederived by using that bracket of
the Miura variable (free-field) $\phi$.
Permuting the Miura variable ($\phi\to -\phi$) gives a new Lax operator
 $L_\Om=p^2+u_\Om=(p+\phi)\star(p-\phi)$ with $u_\Om(\phi)=u(\phi)-4\ka\phi'$.
Now suppose the solutions of the Moyal KdV equations can be parametrized by a single
  function  $\tau$, the  so-called tau-function, such that $u(x;t)=8\ka^2\pa_x^2\ln\tau(x;t)$.
  Then the Adler's approach to the B\"acklund transformation leads to the following
    transformation rule for tau functions
\be
\tau(x;t)\to \tau_\Om(x;t)=\exp\left[-\frac{1}{2\ka}\int^x\phi\right]\cdot\tau(x;t)
\label{tau}
\ee
which reduces to the ordinary case\cite{CSY} when $\ka=1/2$.

In summary, we have investigated the B\"acklund transformation for the Moyal KdV hierarchy
from the viewpoint of KW theorem. It turns out that the cyclic permutations of the Miura
variables provide the elementary B\"acklund transformations of the hierarchy which can
be expressed in terms of tau functions. It would be interesting to see whether
the Moyal KdV hierarchy (\ref{Moyalkdv}) has corresponding Hirota bilinear equations with
 (\ref{tau}) as a symmetry.\\
{\bf Acknowledgments\/}\\
  M.H.T thanks the National Science Council of Taiwan (Grant numbers NSC 89-2112-M-194-020
  and NSC 90-2112-M-194-006)
   for support.

\end{document}